\documentclass[galaxies,article,accept,moreauthors,pdftex,10pt,a4paper]{mdpi}
\firstpage{1}
\articlenumber{x}
\doinum{10.3390/------}
\pubvolume{xx}
\pubyear{2017}
\copyrightyear{2017}
\history{Received: 29 June 2017; Accepted: date; Published: date}

\pdfoutput=1

\Title{High-Energy Polarization: Scientific Potential and Model Predictions}

\Author{Haocheng Zhang $^{1,2}$}

\AuthorNames{Haocheng Zhang}

\address{%
$^{1}$ \quad Department of Physics and Astronomy, University of New Mexico, Albuquerque, NM 87131, USA; hz193909@unm.edu\\
$^{2}$ \quad Theoretical Division, Los Alamos National Laboratory, Los Alamos, NM 87545, USA}

\abstract{Understanding magnetic field strength and morphology is very important for studying astrophysical jets. Polarization signatures have been a standard way to probe the jet magnetic field. Radio and optical polarization monitoring programs have been very successful in studying the space- and time-dependent jet polarization behaviors. A new era is now arriving with high-energy polarimetry. X-ray and $\gamma$-ray polarimetry can probe the most active jet regions with the most efficient particle acceleration. This new opportunity will make a strong impact on our current understanding of jet systems. This paper summarizes the scientific potential and current model predictions for X-ray and $\gamma$-ray polarization of astrophysical jets. In particular, we discuss the advantages of using high-energy polarimetry to constrain several important problems in the jet physics, including the jet radiation mechanisms, particle acceleration mechanisms, and jet kinetic and magnetic energy composition. Here we take blazars as a study case, but the general approach can be similarly applied to other astrophysical jets. We conclude that by comparing combined magnetohydrodynamics (MHD), particle transport, and polarization-dependent radiation transfer simulations with multi-wavelength time-dependent radiation and polarization observations, we will obtain the strongest constraints and the best knowledge of jet physics.}

\keyword{active galaxies; blazars; gamma-ray bursts; radiation mechanisms; polarization; magnetohydrodynamics; particle acceleration}

\conferencetitle{Polarised Emission from Astrophysical Jets}

\begin{document}


\section{Introduction}

Blazars are the most violent class of active galactic nuclei. To understand blazar jet physics, several key processes need to be studied---namely, the magnetic field evolution, the particle acceleration, and the radiation mechanisms. Spectral fitting and multi-wavelength light curves have been used to study the blazar jet physics. However, these methods cannot constrain the magnetic field evolution. Radio to optical polarization measurements have been a standard probe of the jet magnetic field.
In particular, recent observations of $\gamma$-ray flares with optical polarization angle swings and
substantial polarization degree variations indicate the active role of the magnetic field during
flares \cite{Abdo10,Blinov15}. Several models have been put forward to explain
these phenomena \cite{Larionov13,Marscher14}, including first-principle magnetohydrodynamics-based models \citep{ZHC16,ZHC17}. However, radio and optical polarization signatures may often be coming from regions that do not emit strong high-energy radiation, in which case they do not probe the most active particle acceleration regions so as to probe the key jet physics there.

\scalebox{0.96}{Currently, there are several high-energy polarimeters being proposed or under development~\cite{Produit05,Beilicke12,Weisskopf16,Tatischeff16}.} High-energy polarimetry will provide unique insights into the jet physics. The most obvious breakthrough will be a better multi-wavelength coverage. MeV gamma-ray polarimeters will bridge the MeV gap in the blazar spectrum so that we can obtain light curves and polarization from radio all the way to gamma rays. Second, while the radio to optical blazar emission is believed to be electron-produced synchrotron, the X-ray and gamma-ray emission is more complicated. Leptonic models suggest that it can come from Compton scattering of synchrotron emission or external emission, but hadronic models argue for proton-produced synchrotron and hadronic cascades. High-energy polarimetry can distinguish between synchrotron and Compton scattering through the study of the polarization degree. Third, blazars show multi-wavelength variations, indicating strong particle acceleration. Generally, gamma-ray energies exhibit the strongest flares. This implies that the high-energy emission comes from the most active acceleration regions with the most energetic particles. Both shock and magnetic reconnection scenarios may be responsible for the particle acceleration, but they have different magnetic field evolution, which can be diagnosed with high-energy polarimetry. Finally, current blazar jet models suggest that the jet is highly magnetized close to the central engine~\cite{Tchekhovskoy11}. At the so-called blazar zone around pc to tens of pc, a large amount of jet energy converts to multi-wavelength emission. Then, the jet continues to hundreds of kpc or even Mpc. Two major questions emerge: first, after the jet launches, is it dominated by kinetic energy or magnetic energy, or is it in equipartition? Second, is the blazar zone the place where most of the magnetic energy dissipates? High-energy polarimetry can probe the jet magnetic field, especially in the blazar zone. To understand these issues, we need to lay out robust physical modeling. In particular, we need to treat magnetic field and particle evolution carefully in order to obtain the correct radiation and polarization signatures. In the following, we will discuss the averaged polarization degree from X-ray to $\gamma$-ray based on steady-state spectral fitting in Section \ref{sec2}
, time-dependent polarization signatures of shock and magnetic reconnection scenarios based on detailed particle evolution and polarization-dependent radiation transfer in Section \ref{sec3}, and general trends of polarization signatures in the kinetic-dominated and magnetic-dominated jet emission environment, based on magnetohydrodynamics (MHD)-integrated modeling of polarization signatures.

\section{High-Energy Polarization Degree of Leptonic and Hadronic Blazar Models}\label{sec2}

Based on the peak frequency of the low-energy component, blazars can be classified into low-synchrotron-peaked (LSP, peaked around infrared), intermediate-synchrotron-peaked (ISP, peaked around optical to ultraviolet), and high-synchrotron-peaked (HSP, peaked at X-ray) blazars.
While the low-energy component of blazar spectra is generally believed to be
synchrotron radiation of ultrarelativistic electrons, the origin of the high-energy component is still unclear. The leptonic model argues that the high-energy component is due to
inverse Compton scattering by nonthermal electrons of either the electron synchrotron emission \cite{Marscher85,Maraschi92}
or external photon fields \cite{Dermer92,Sikora94}. On the other hand, the hadronic model suggests that the
high-energy emission is dominated by synchrotron emission of ultrarelativistic protons and the cascading secondary
particles due to hadronic processes \cite{Mannheim92,Mucke01}. Both models
are able to produce reasonable fits to snap-shot SEDs
~of blazars \cite{Boettcher13}. Therefore,
additional diagnostics are necessary.

An obvious diagnostic is the radiation mechanism. It is well-known that linear polarization
arises from synchrotron radiation of relativistic charged particles in ordered
magnetic fields, while Compton scattering off
~relativistic electrons will reduce
the degree of polarization of the target photon field, without entirely destroying
it. Compton scattering of unpolarized target photon fields by isotropic distributions
of electrons (and positrons) will always result in un-polarized Compton emission.
The well-known formalism for calculating synchrotron polarization can be found
in textbooks \cite{RL85}. An analytical formalism for evaluating the polarization of
Compton-scattered radiation in the Thomson regime was developed by \cite{BCS70}
and applied specifically to synchrotron self-Compton (SSC) emission by \cite{BS73}.
More recently, \cite{Krawczynski12} provided a general Monte-Carlo-based framework
for the evaluation of polarization signatures in relativistic environments in both
Thomson and Klein--Nishina regimes, verifying that the expressions of \cite{BCS70}
and \cite{BS73} are valid in the Thomson regime. However, these calculations generally assume a purely power-law distribution of electrons.

\begin{figure}[H]
\centering
\includegraphics[width=0.45\linewidth]{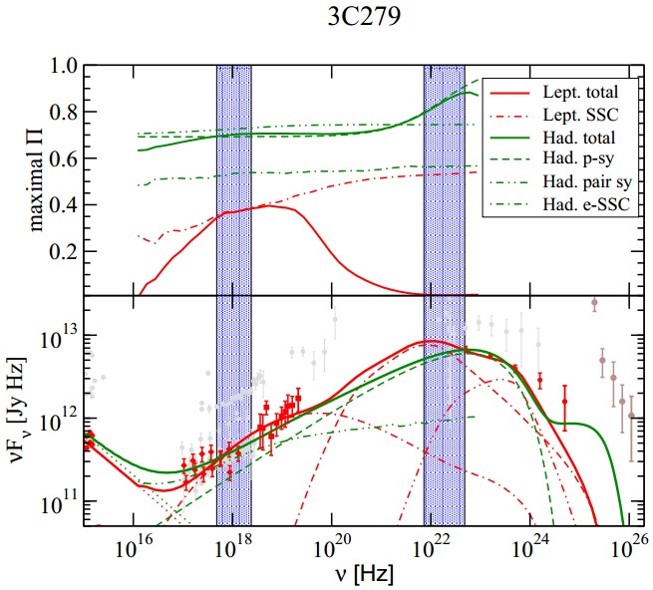}
\includegraphics[width=0.45\linewidth]{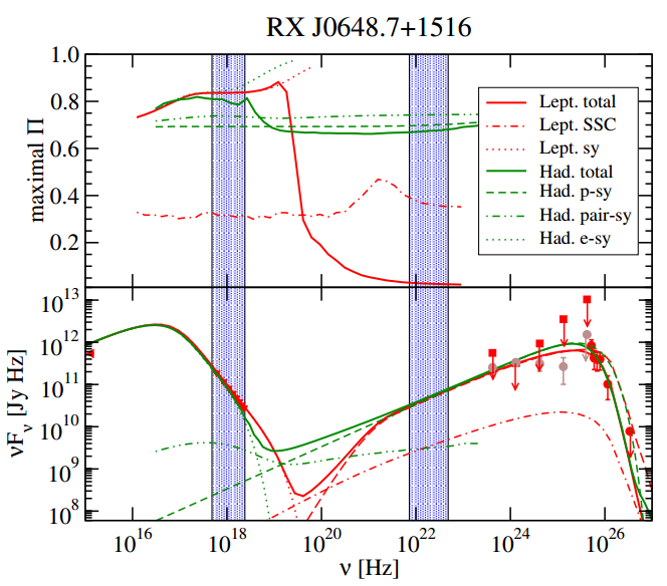}
\caption{UV through $\gamma$-ray SEDs
~(lower panels) and the corresponding maximum degree of
polarization (upper panels) for the two blazars 3C279 (\textbf{left}) and RX J0648.7+1516
(\textbf{right}). Leptonic model fits are plotted in red, hadronic models in green.
Different line styles indicate individual radiation components, as labeled
in the legend. Shaded areas indicate the 2--10~keV X-ray range
and the 30--200~MeV range. The polarization degree calculation assumes a perfectly ordered magnetic field. This figure is reproduced from \cite{ZHC13}. SCC: synchrotron self-Compton.}
\label{poldeg}
\end{figure}

We have calculated the maximal X-ray to $\gamma$-ray polarization degree for both leptonic and hadronic one-zone blazar models \cite{ZHC13}, based on detailed spectral fitting \cite{Boettcher13}. At a steady state, we expect very different polarization degrees for leptonic and hadronic models. For flat spectrum radio quasars such as 3C279, we can see that the keV to MeV polarization degree differs significantly for the two scenarios (Figure \ref{poldeg}). However, for high-frequency-peaked BL Lacs, since the X-ray emission is dominated by the electron synchrotron, we see no difference in the keV band polarization (Figure \ref{poldeg})---in MeV the polarization degree is still very different. Therefore, we expect that high-energy polarimeters---in particular MeV gamma-ray polarimeters---can easily distinguish the leptonic and hadronic models.

However, these calculations are based on a perfectly ordered magnetic field. In reality, optical polarimetry has shown that the blazar magnetic field is only partially ordered. Since the low-energy and high-energy components often show simultaneous flares, we expect that they are coming from the same region. Therefore, we can use optical polarization to determine how ordered the magnetic field is. Generally, the optical polarization degree ranges between $10-20\%$ for a maximal polarization degree of  about $70\%$ (similar to the proton synchrotron), so we expect that the hadronic model will also result in $10-20\%$ polarization. On the other hand, the leptonic model predicts a polarization degree up to $40\%$, then in the usual situation, less than $10\%$. { Chakraborty et al.}~\cite{Chakraborty15} have shown that current-generation high-energy polarimeters can generally detect $10\%$ polarization in bright blazars. Thus, we expect that X-ray polarimeters may have a chance to diagnose leptonic and hadronic models for LSPs during strong X-ray flares for some very active blazars (e.g., 3C~279). However, since the X-ray emission of ISPs and HSPs includes considerable contributions from low-energy synchrotron component, X-ray polarimeters are unlikely to distinguish the two scenarios for ISPs and HSPs. On the other hand, gamma-ray polarimeters can distinguish the two scenarios for most blazars.

\section{Time-Dependent High-Energy Polarization of Shock and Magnetic Reconnection}\label{sec3}

Blazars exhibit strong multi-wavelength variability \cite{Aharonian07}, indicating active particle acceleration. Both shock and magnetic reconnection scenarios may be responsible for the particle acceleration. Shock models generally assume that the emission region has a significant amount of kinetic energy, which can be converted to nonthermal particle energy through the shock acceleration \cite{Achterberg01,Spitkovsky08}. On the other hand, particle-in-cell (PIC) simulations have shown that magnetic reconnection can also make power-law nonthermal particle spectra \cite{Sironi14,Guo14}. In particular, in a proton-electron plasma, PIC simulations have shown that magnetic reconnection can accelerate both protons and electrons \cite{Guo16}. Both mechanisms can produce power-law shaped particle spectra, which are consistent with blazar spectral fitting \cite{Boettcher13}. However, the two mechanisms have different magnetic field evolution, which can be diagnosed by polarization signatures.

\begin{figure}[H]
\centering
\includegraphics[width=0.55\linewidth]{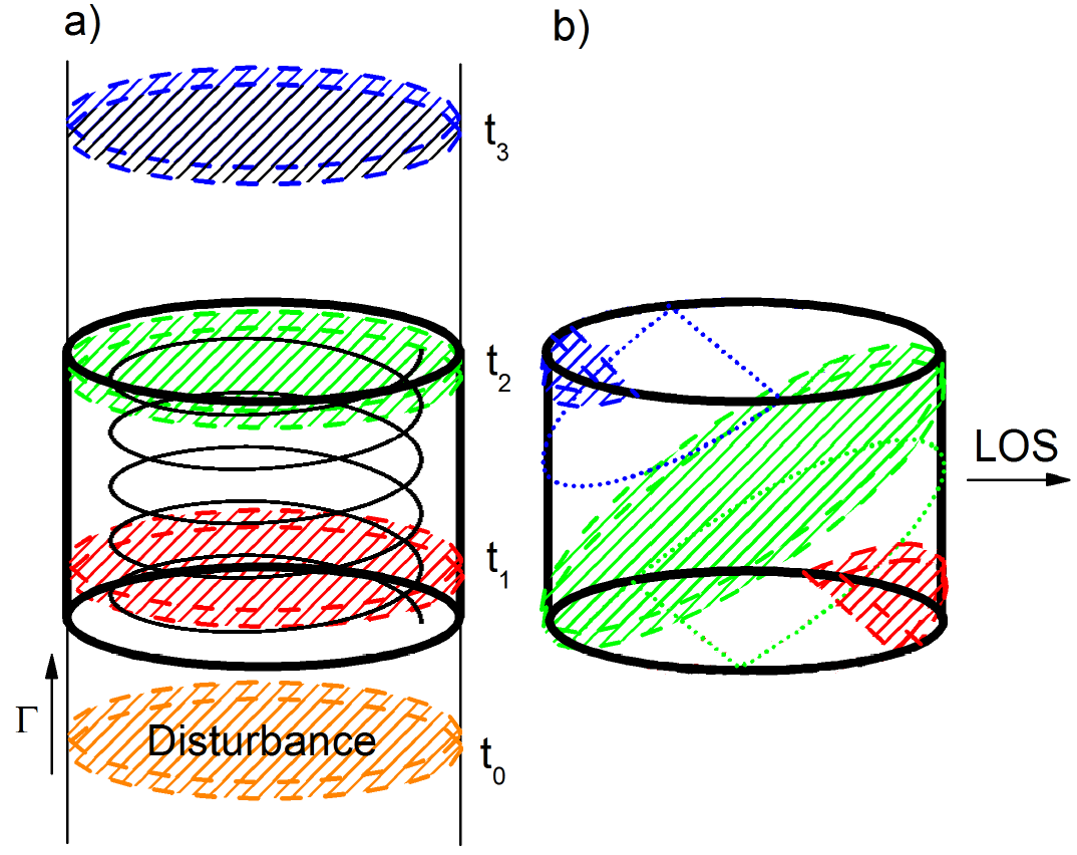}
\includegraphics[width=0.35\linewidth]{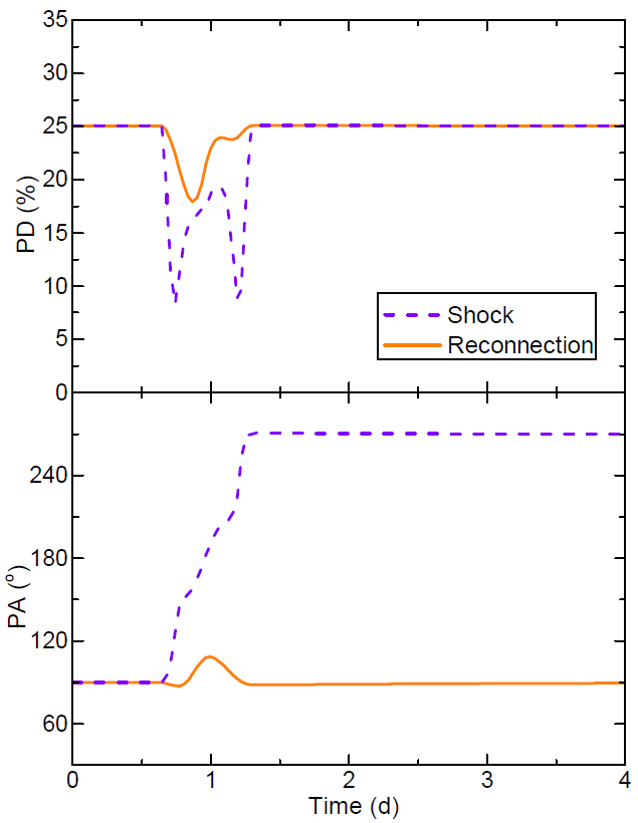}
\caption{\textbf{Left}: sketch of the model and light crossing time. \textbf{a}) a disturbance propagates through the emission region pervaded
by a helical magnetic field in its comoving frame. Red, green, and blue colors denote the location of the
disturbance at approximately entering ($t_1$), leaving the emission region ($t_2$), and some time after
leaving the emission region ($t_3$). \textbf{b}) the corresponding flaring region at the $t_1$ to $t_3$ at equal
photon-arrival times at the observer. Dashed shaded regions are the active region; the region between the
dashed shaded region and the dotted shape is the evolving region. \textbf{Right}: time-dependent polarization signatures of shock and magnetic reconnection scenarios. Upper panel is the polarization degree, and the lower panel is the polarization angle. The calculation is performed by combining particle evolution code (3DHad) and the polarization-dependent radiation transfer code (3DPol). $90^{\circ}$ polarization angle is defined to be along the jet axis. The figure
~is reproduced from \cite{ZHC16b}.}
\label{ltte}
\end{figure}

Recent observations of simultaneous polarization angle swings and multi-wavelength flares demonstrate that the magnetic field is actively evolving along with particle acceleration \cite{Abdo10,Blinov15,Blinov16}. These phenomena are expected from both shock and magnetic reconnection scenarios. In order to distinguish the two scenarios, we need to extend the modeling for variability. In this situation, the light crossing and particle cooling need to be considered. As an illustration of light crossing time scale, consider the case that a flat shock propagates through an emission region, and we are observing to the right. Since the emission from the left side will arrive later than that from the right side, the observed flaring region that is perturbed by the shock is distorted into an elliptical shape (Figure \ref{ltte}). In this case, the polarization signatures will first represent the magnetic field on the right, then the averaged magnetic field across the emission region, and finally the magnetic field on the left. If the emission region is pervaded by a helical magnetic field, since the magnetic field directions are opposite from the right to the left, this light crossing effect may naturally give a polarization change. However, this also depends on particle cooling. If the particle cooling time scale is significantly shorter than the light crossing time scale, then we observe nonthermal particles moving from the right to the left, following the shock perturbation. However, if the cooling time scale is comparable to or even longer than the light crossing time scale, then the nonthermal particles may occupy a much larger part of the emission region than the shock perturbed part, which smooths out the difference of magnetic field structure from the right to the left. In this case, we do not expect a strong change in the polarization.

In order to study these two effects, we need to consider both particle transport and polarization-dependent radiation transfer. For the particle transport, PIC simulation can track both particle and magnetic field evolution under first principles \cite{Spitkovsky08,Sironi14,Guo14}. However, it is very computationally expensive, and so far it cannot be directly applied to study real blazar observations. For practical uses, a good approach is to solve a Fokker--Planck equation for the particles \cite{Stawarz08,Chen14,Diltz14,Weidinger15}. This usually has low computational cost, and provides a good approximation of particle evolution. Since polarization is a 3D effect, and the magnetic field as well as particle distributions are unlikely to be homogeneous across the emission region, we generally need to solve Fokker--Planck equations in an inhomogeneous 3D emission region. For the polarization-dependent radiation transfer, the polarization-dependent radiation transfer code (3DPol) can be easily coupled with magnetohydrodynamics or particle transport simulations \cite{ZHC14}. It allows inhomogeneous magnetic field and particle distributions, and includes detailed polarization-dependent radiation transfer, with time-, space-, and frequency-dependencies.

\begin{figure}[H]
\centering
\includegraphics[width=0.6\linewidth]{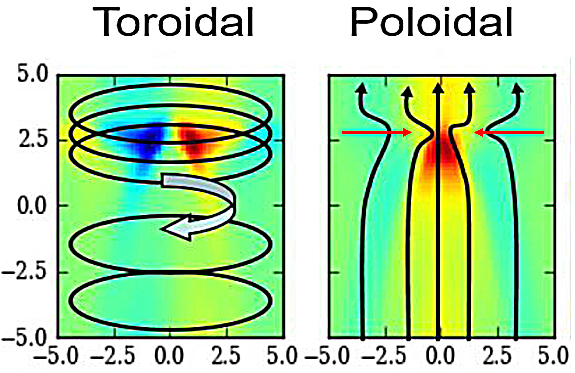}
\includegraphics[width=0.6\linewidth]{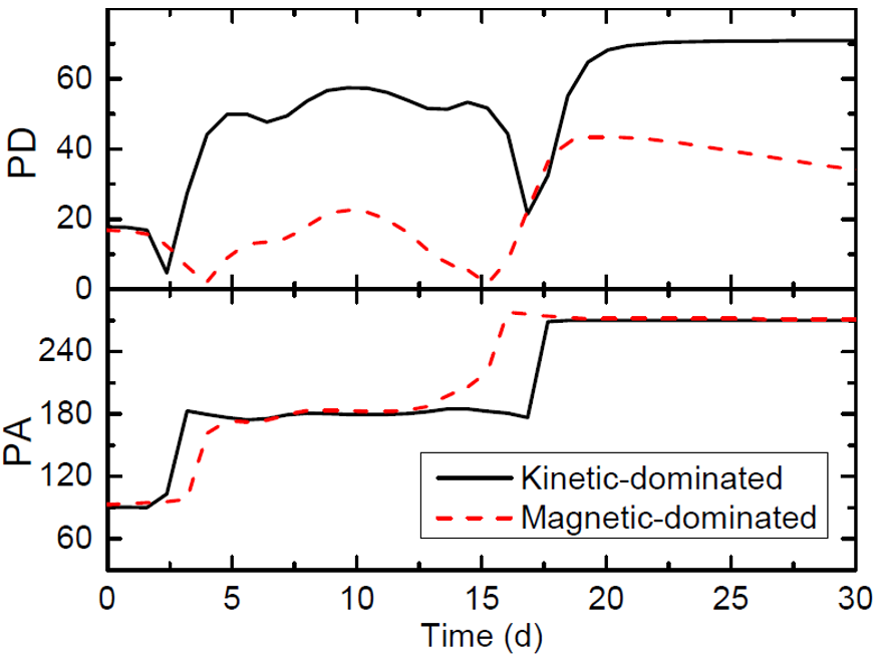}
\caption{\textbf{Top}: sketch and simulation results of the toroidal and poloidal magnetic field components under shock compression. Left panel shows that the toroidal component is compressed at the shock front. Right panel shows the toroidal component is squeezed by the excess magnetic force due to the toroidal magnetic field compression. \textbf{Bottom}: time-dependent polarization signatures of kinetic-dominated and magnetic-dominated jet emission environment. Upper panel is the polarization degree and the lower panel is the polarization angle. The calculation is based on combining magnetohydrodynamics (MHD) simulation (LA-COMPASS) with polarization-dependent radiation transfer (3DPol). Figure is reproduced from \cite{ZHC16}.}
\label{shock}
\end{figure}

These two effects can potentially help us to distinguish between the shock and magnetic reconnection scenarios. In the case of a shock, the shock will compress the magnetic field, leading to faster particle cooling during flares than the quiescent state. Therefore, the flare may be accompanied by a major polarization variation (Figure \ref{ltte}). On the other hand, magnetic reconnection will dissipate the magnetic field, so that the particle cooling is slower than the quiescent state. If the particle cooling becomes comparable to the light crossing time, then we do not expect any strong polarization change during flares. This is particularly important for the hadronic model, since the proton cooling time scale is generally comparable to the light crossing time scale in the quiescent state. During the flaring state, the particle cooling time scale can significantly shift away from the quiescent state, which can be tracked by the particle evolution simulations. By coupling nonthermal particle evolution with polarized radiation transfer, we can distinguish the two scenarios with future high-energy polarimetry.


\section{Polarization Signatures of Kinetic-Dominated and Magnetic-Dominated Jets}

Another major effect for blazar variability is the magnetic field evolution, which is strongly dependent on how much the jet is magnetized. For example, in the case of a shock propagating through the emission region (Figure \ref{shock}), in a jet dominated by kinetic energy, the shock compression at the shock front can be very strong, while the magnetic force is relatively weak. This will lead to a strong change in the magnetic strength and topology inside the emission region. Moreover, since the magnetic field is strongly perturbed by the shock, it takes a long time to restore its initial state. On the other hand, in a magnetic-dominated jet, the shock is relatively weak. Thus, we do not expect strong change in the magnetic field during the shock perturbation, and the magnetic field can quickly recover its initial status.

This provides a great chance to couple MHD simulations with polarized radiation transfer, so that we can track magnetic field evolution under first principles \cite{Mizuno09,Guan14,Deng16}. Indeed, for a kinetic-dominated jet, due to the strong shock compression (Figure \ref{shock}), the magnetic field becomes highly ordered at the shock front. This results in a strong change in the polarization signatures---in particular, a very high polarization degree during flares. Additionally, the system cannot restore its initially partially-ordered magnetic field after the shock perturbation, and thus the polarization degree is still very high. However, for a magnetic-dominated jet, the change in the polarization is much smaller during flare, and we observe a clear restoration phase of the polarization signatures at the end of the flare. From optical polarization monitoring, we know that the polarization degree cannot be too high (mostly around $10\%$), and if a major change in the polarization happens, it will quickly go back to the quiescent state after the change \cite{Abdo10,Blinov15}. This suggests that the optical polarization signatures prefer a magnetized jet environment. However, optical polarization can come from a larger region that may contaminate the polarization signatures in the most active particle acceleration regions. High-energy polarimetry will probe the most active regions to constrain how much the jet is magnetized.


\vspace{6pt}


\acknowledgments{H.Z. thanks the anonymous referee for valuable suggestions. This publication has received funding from the European Union’s Horizon 2020 research and innovation programme under grant agreement No 730562 [RadioNet]. H.Z. acknowledges the support of AAS International Travel Grant for attending this conference. This work is is supported by the LANL/LDRD program and by DoE/Office of Fusion Energy Science through CMSO. Simulations were conducted on LANL's Institutional Computing machines.}

\end{document}